\documentclass[twocolumn,showpacs,prl]{revtex4}
\usepackage{graphicx}
\usepackage{latexsym}

\begin{document}

\title{Field-Effect Persistent Photoconductivity}

\author{E.\ P.\ De Poortere, Y.\ P.\ Shkolnikov, and M.\ Shayegan}
\address{Department of Electrical Engineering, Princeton University,
Princeton, New Jersey 08544}

\date{\today}

\begin{abstract}
We report a persistent increase or decrease in the two-dimensional electron density of AlAs or GaAs quantum wells
flanked by AlGaAs barriers, brought about by illuminating the samples at $T \sim 4$ K while simultaneously
applying a voltage bias between a back gate and the two-dimensional electron gas. Control of the final carrier
density is achieved by tuning the back gate bias {\it during} illumination. Furthermore, the strength of the
persistent photoconductivity depends on the Al mole fraction in the back Al$_{x}$Ga$_{1-x}$As barrier, and is
largest at $x \simeq 0.4$.
\end{abstract}


\maketitle

Several compound semiconductors, when illuminated briefly at low temperatures ($T \lesssim 100$ K) with infrared
or visible light, retain their photoconductivity for times that vary from a few minutes to hours or days. This
remarkable phenomenon, called {\it persistent photoconductivity} (PPC), has been mostly observed in doped II-IV
and III-V semiconductors such as AlGaAs:Si/Te \cite{lang79}, ZnCdTe:Cl \cite{burkey76}, GaAsP:Te/S
\cite{craford68}, and AlN:Si \cite{zeisel00}, in which it likely results from the formation of ``{\it
DX}-centers''. The latter are charged defect centers that behave as deep donors, and transform into metastable
shallow donors under appropriate illumination. A successful microscopic model for the {\it DX}-centers in AlGaAs,
based on large lattice relaxation, has been established by Chadi and Chang \cite{chadi89}.

A general picture of PPC, however, is still conspicuously missing, and alternative models are still being invoked
to explain its origin. Indeed, PPC also occurs in materials that do not contain {\it DX}-centers
\cite{theis86,reddy98,chen98,zervos99}, and can actually result in a {\it reduction} of the electrical
conductivity, in which case it is referred to as {\it negative} PPC \cite{chou85,pettersson93}. Aside from the
{\it DX}-center model, another mechanism put forward to account for PPC involves the photoexcitation and
subsequent separation of electron-hole pairs, followed by trapping of some of the electrons (or holes) by the
spacers/barriers \cite{theis86,zervos99}. Deep levels in undoped materials such as GaN and GaAs may also result
from the formation of anion antisites (e.g., N$_{Ga}$) or vacancies (V$_{Ga}$ or V$_{As}$ in GaAs)
\cite{reddy98,chen98}. In many cases, PPC probably stems from the cumulative effect of electron-hole pair
excitation and relaxation of defects similar to the {\it DX}-centers.

\begin{figure}
\includegraphics[scale=.8]{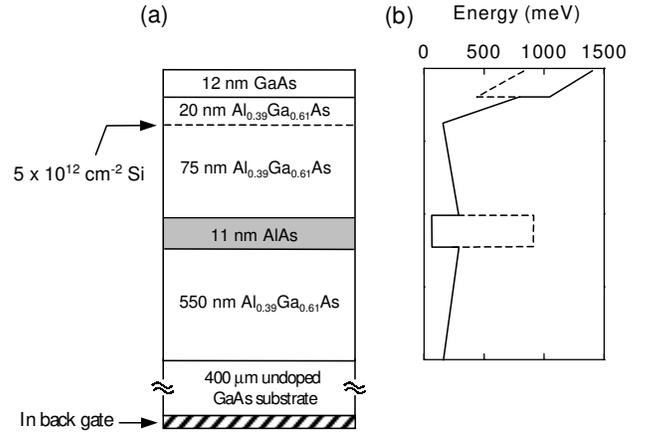}
\caption{(a) Layer structure of the AlAs QW. (b) Schematic conduction band diagram of the sample. X-point and
$\Gamma$-point conduction band edges are indicated by solid and dashed lines, respectively. } \label{structure}
\end{figure}

The main result we wish to describe here is that the two-dimensional (2D) carrier density ($n$) in a GaAs or AlAs
quantum well (QW) with suitable Al$_{x}$Ga$_{1-x}$As barriers ($x \sim 0.4$), can be tuned reversibly from almost
zero to values larger than 5 $\times 10^{11}$ cm$^{-2}$, by applying a small electric field ($< 500$ V/cm) between
the 2D electrons and a back gate, while briefly illuminating the sample at $T \simeq 4$ K with red LED light. In
other words, the zero-frequency dielectric permittivity of Al$_{0.39}$Ga$_{0.61}$As, which is equal to
10$\epsilon_0$, increases effectively by two orders of magnitude when the material is illuminated at low
temperatures, as calculated from the geometric capacitance of the sample. In samples with Al$_{0.39}$Ga$_{0.61}$As
barriers, the carrier density thus induced in the QW remains approximately constant after light is turned off,
while with barriers with a lower Al content, the photoconductivity is only partly persistent. Though most of our
measurements were done in AlAs QWs, we also have some limited data on GaAs QWs, which we will discuss later in
this report.

To quantify this field-effect PPC (FEPPC), we have performed measurements in an 11 nm-wide AlAs QW grown by
molecular beam epitaxy (MBE), surrounded by Al$_{0.39}$Ga$_{0.61}$As barriers, with a single Si front dopant layer
separated from the QW by a 75 nm-thick Al$_{0.39}$Ga$_{0.61}$As spacer (Fig.\ \ref{structure}). Two samples were
cut from the same wafer, one unpatterned and contacted in the van der Pauw geometry (sample A), and one patterned
as a Hall bar mesa (sample B). Both were fitted with a gate located on the back of the 400 $\mu$m-thick, undoped
GaAs substrate. 150 nm-thick AuGeNi contacts were deposited and alloyed at 440 $^o$C, and samples were cooled in
the dark, either in a $^4$He dewar, or in a $^3$He cryostat kept at $T \sim 4$ K. A red LED (wavelength 660 nm)
was placed next to the sample at a distance of about 1 cm. The carrier density was determined through measurements
of the Hall resistance ($R_{xy}$). From transport measurements done on this and other samples, we have confirmed
that the density deduced from $R_{xy}$ at 4 K is the same as the 2D carrier density obtained from Shubnikov-de
Haas data at 0.3 K. Magnetotransport data in AlAs samples illuminated using the technique described here have been
reported elsewhere \cite{depoortere02}.

\begin{figure}
\includegraphics[scale=0.4]{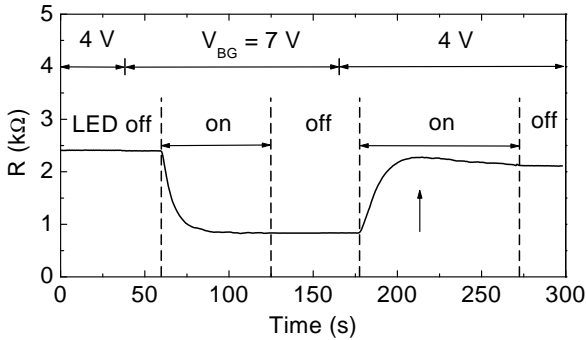}
\caption{Transient of the four-point resistance ($R$) of 2D electrons in an AlAs QW bordered by
Al$_{0.39}$Ga$_{0.61}$As barriers, as the sample is illuminated with a red LED, while a voltage bias ($V_{BG}$) is
applied between the 2D electrons and a back gate (sample A). $R$ first {\it decreases} as the LED is turned on at
a higher $V_{BG}$, while the resistance {\it increases} when light is applied while $V_{BG}$ is reset to its
original value of 4 V. The peak value reached by $R$ during this second illumination is about 5 \% lower that its
value at $t = 0$. $R$ begins to decrease slowly after reaching its maximum, at a time marked by the vertical
arrow.} \label{LEDon}
\end{figure}
Figure \ref{LEDon} illustrates the effect of illumination on the four-point resistance ($R$) of sample A with an
applied back gate bias ($V_{BG}$), at $T = 4.2$ K. At time $t = 0$, $V_{BG} = 4$ V \cite{initialR}. $V_{BG}$ is
first raised to 7 V, which changes $R$ by a negligible amount in the dark. As the LED is turned on, $R$ drops from
2400 to 830 $\Omega$ in about 30 s, after which it remains constant. The LED current is then turned off at $t =
125$ s, which does not affect $R$. At $t = 165$ s, $V_{BG}$ is set back to 4 V, which does not change $R$ in the
dark, and then at $t = 177$ s, the LED is turned back on. As a result, $R$ now {\it increases} to reach a maximum
of 2270 $\Omega$, slightly lower than its value at $t = 0$. After reaching this maximum, $R$ slowly decreases
while the LED is on. Finally, the LED is turned off again at $t = 270$ s, keeping $R$ unchanged. The data in Fig.\
\ref{LEDon} thus show that the resistance drop obtained after illumination at $V_{BG} = 7$ V can be partly {\it
reversed} by illuminating the sample at the ``original'' $V_{BG} = 4$ V. The initial $R = 2400$ $\Omega$ cannot be
fully recovered at $V_{BG} = 4$ V, however, indicating that the sample keeps some memory of its illumination. This
initial $R$ can nevertheless be retrieved if the LED is turned on at $V_{BG}$ {\it lower} than 4 V.

\begin{figure}
\includegraphics[scale=0.4]{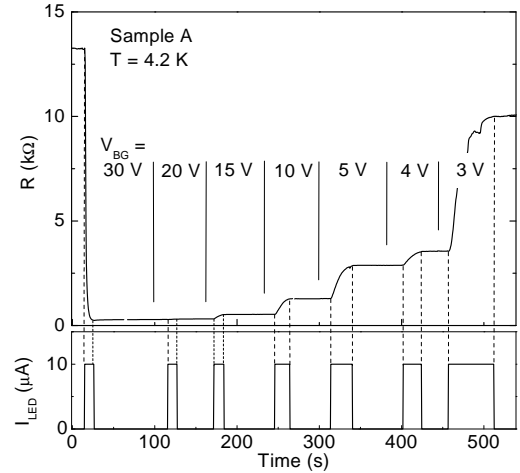}
\caption{Time dependence of the four-point resistance of 2D electrons in an AlAs QW, following the illumination
sequence shown in the lower panel. Each illumination takes place at a different back gate bias, and is stopped
when $R$ reaches its maximum value (vertical arrow in Fig.\ \ref{LEDon}).} \label{timedep}
\end{figure}

The evolution of $R$ as we illuminate sample A at successively lower $V_{BG}$ is plotted in Fig.\ \ref{timedep},
obtained in a separate run. The current passing through the LED is plotted in the lower panel as a function of
time. At time $t = 0$, the sample density is low ($< 2 \times 10^{11}$ cm$^{-2}$), and $R$ is correspondingly
high. At $t = 15$ s, we illuminate the sample at $V_{BG} = 30$ V until $R$ drops to 260 $\Omega$ and stops
decreasing. We then reduce $V_{BG}$ to 20 V, which does not affect $R$ (even as we shine light upon the sample).
We further decrease $V_{BG}$ to 15 V, which in the dark does not change $R$ either. The LED is then turned on: $R$
increases, quickly reaches a maximum value of 530 $\Omega$, then starts to decrease slowly, similarly to the
behavior seen in Fig.\ \ref{LEDon}. We turn off the LED right after $R$ reaches its maximum value (For our
experiment at lower $V_{BG}$, this time is marked by a vertical arrow in Fig.\ \ref{LEDon}). $R$ remains constant
after illumination has stopped. This procedure is then repeated by (1) lowering $V_{BG}$, (2) illuminating the
sample, thereby increasing $R$, and (3) turning the LED off as $R$ comes close to its peak value. From the data of
Fig.\ 3, we see that the resistance obtained after illumination increases as $V_{BG}$ decreases. When light is
applied at $V_{BG}$ lower than 3 V, our contact resistance to the 2D electrons becomes too high for $R$ to be
measured reliably.

\begin{figure}
\centering
\includegraphics[scale=.34]{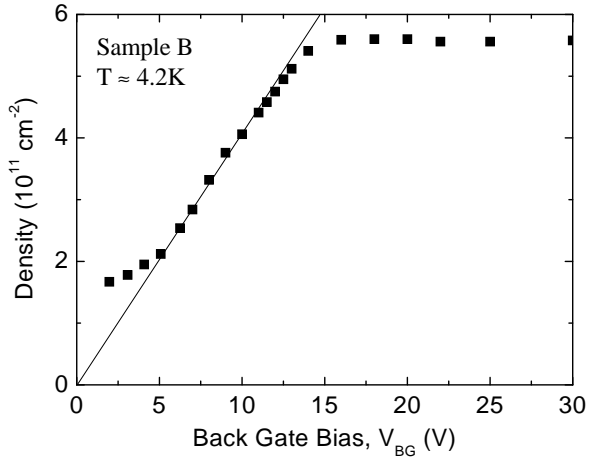}
\caption{2D carrier density in an AlAs QW after illumination at gate bias $V_{BG}$, as a function of $V_{BG}$.
Starting from a high density $n = 5.6 \times 10^{11}$ cm$^{-2}$, $n$ is lowered by illuminating the sample at
gradually lower $V_{BG}$, as in Fig.\ \ref{timedep}.} \label{FEPPCg}
\end{figure}

A similar illumination sequence (with a denser set of back gate biases) was applied to sample B, and the carrier
density was measured after every illumination. The density obtained after each exposure to light at $V_{BG}$ is
plotted as a function of $V_{BG}$ in Fig.\ \ref{FEPPCg}. For $V_{BG} > 16$ V, the density remains constant at $n
\simeq 5.6 \times 10^{11}$ cm$^{-2}$, while for lower biases ($5 < V_{BG} < 14$ V), $n$ decreases quickly and
approximately linearly with $V_{BG}$. It is worth noting that the change in $n$ with $V_{BG}$ is about 200 times
larger than what it would be if this experiment were realized without illumination: in this case, the slope of
$n(V_{BG})$ would be about $2 \times 10^{8}$ cm$^{-2}$/V \cite{BGleak}.

For $V_{BG} \lesssim 5$ V, the density decreases more slowly as $V_{BG}$ is lowered. We note that for $n$ lower
than $\sim 1.7 \times 10^{11}$ cm$^{-2}$, our contact resistance in this sample becomes prohibitively high,
preventing an accurate measurement of the density. As indicated earlier for sample A, we also note that the
densities plotted in Fig.\ \ref{FEPPCg} depend on the history of conditions applied to the sample during a given
cooldown: a repeat of the sequence of illuminations at incremental values of $V_{BG}$ tends to shift the curve of
Fig.\ \ref{FEPPCg} to lower $V_{BG}$.

All other AlAs QWs with Al$_{0.39}$Ga$_{0.61}$As barriers we have measured so far display a behavior qualitatively
similar to that of Figs.\ \ref{LEDon} - \ref{FEPPCg}: $n$ can be tuned controllably from less than 2 to about $5
\times 10^{11}$ cm$^{-2}$, by illuminating the sample at $0 < V_{BG} < 20$ V. In QWs confined within lower-x
Al$_{x}$Ga$_{1-x}$As barriers, however, the maximum density obtained after illumination is lower than in samples
with $x \simeq 0.4$. We define this ``maximum'' density ($n_{max}$) in a given sample as that reached by
increasing $V_{BG}$ and by illuminating the sample at $T \sim 4$ K for brief intervals ($< 5$ seconds) until $n$
saturates. The density obtained in this manner is characteristic of the sample, in the sense that it does not
depend on cooldown or on the history of illumination and gating \cite{longillum}.

\begin{figure}
\centering
\includegraphics[scale=.4]{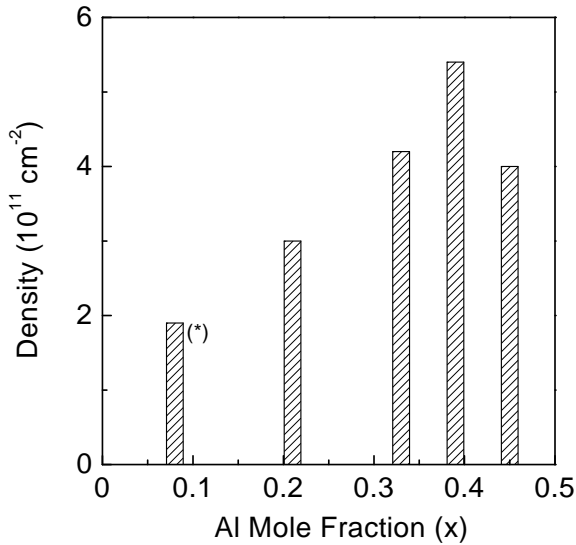}
\caption{Dependence of the electron density in an AlAs quantum well on the Al concentration ($x_B$) of back AlGaAs
barriers. In each sample, a 2D electron gas is obtained after illumination at positive $V_{BG}$, as explained in
the text. The bars indicate the range of 2D densities that can be obtained in the corresponding samples. The
highest densities occur for $x_B \simeq 0.4$. Density data obtained for $x_B = 0.4$ and $x_B = 0.45$ were repeated
in several wafers with nominally the same Al content in the back barrier: four wafers with $x_B = 0.4$, and three
different wafers with $x_B = 0.45$. All sample structures are based on the layout shown in Fig. \ref{structure}.
(*) Because of the low Al content ($x_B = 0.08$) of the back barrier in this sample, 2D electrons are located both
in the AlAs QW and in the Al$_{0.08}$Ga$_{0.92}$As.} \label{Almolefraction}
\end{figure}

In Fig.\ \ref{Almolefraction}, we plot $n_{max}$ for different AlAs samples, as a function of the Al
concentrations ($x_B$) in the back AlGaAs barrier. The structures grown above the AlAs QW in these samples are all
similar to the one shown in Fig.\ \ref{structure}.  As $x_B$ decreases from 0.39 to 0.08, $n_{max}$ decreases from
about 5.6 to 1.9 $\times 10^{11}$ cm$^{-2}$, showing a direct correlation between $x_B$ and sample density.
Furthermore, out of the several samples grown with $x_B$ = 0.45, none has a density higher than $4.0 \times
10^{11}$ cm$^{-2}$, indicating that the strongest FEPPC is obtained with $x_B \simeq 0.4$. We point out that, when
$x \gtrsim 0.4$, the forbidden gap in AlGaAs becomes indirect, so that photons indeed are much less efficiently
absorbed by the barriers.

Although we do not have a quantitative explanation for the phenomenon described in this report, we can
nevertheless draw a qualitative picture of FEPPC from our measurements. First, since the effect depends
sensitively on the Al concentration in the AlGaAs barrier {\it underneath} the QW, we suggest that the charged
centers responsible for the increase in 2D carrier density after illumination are also located below the QW
\cite{GaAsPPC}. In addition, because of screening, the back gate has little effect on electric fields above the QW
when 2D electrons are present in the QW. Thus, since the increase in 2D density brought about by illumination at
$V_{BG} > 15$ V is at least $4 \times 10^{11}$ cm$^{-2}$ (Fig.\ \ref{FEPPCg}), we deduce that FEPPC creates
$\gtrsim 4 \times 10^{11}$ cm$^{-2}$ positive charges {\it in the back AlGaAs barrier}.

Our next question concerns the nature of these positive charges in the back AlGaAs layer: can they result from
unintentional impurities present in the MBE during growth, or do they originate from some other kind of crystal
defect? Because contaminants (mainly C) in our MBE are mostly incorporated as {\it acceptors} in the barriers,
they cannot cause the effect we observe. Furthermore, the concentration of unintentional dopants required to
produce a 2D density observed in our measurements needs to be greater than 5 $\times 10^{16}$ cm$^{-3}$, a value
about 500 times larger than the estimated background impurity concentration in our samples. Thus residual
impurities cannot explain the magnitude of the field-and-light-induced electric charge, and we are led to conclude
that this charge results from the presence of crystal defects in AlGaAs, which are able to bind a positive charge
at low temperatures. High quality AlGaAs alloys are notoriously difficult to grow by MBE, so a larger density of
crystal defects is actually expected in this material.

A simple mechanism for FEPPC for $V_{BG} > 0$, outlined below,
can have two possible starting points: either photons absorbed
by the back AlGaAs layer produce electron-hole pairs, which are
then separated by the electric field; or light can induce a
deep-to-shallow transition in levels associated with AlGaAs
defects, thereby generating an effective ($DX$-like) donor in
the barrier. In both scenarios, the field separates positive and
negative charges spatially, attracting electrons towards the
back gate, and repelling the positive charge towards the QW.
Once the LED is turned off, some of the positive charges remain
trapped in the AlGaAs close to the QW, creating an electric
field that is much stronger than the field resulting from
$V_{BG}$. We do not know at this point the nature and energetics
of the charge-trapping defects, or the precise mechanism for
charge transport through the AlGaAs during and after
illumination. The reason why FEPPC is strongest for $x_B = 0.4$
is also unknown, though it could be related to the
near-degeneracy of $\Gamma$ and $X$ conduction band minima in
Al$_{0.4}$Ga$_{0.6}$As \cite{mooney90}. Measurements of the
photoluminescence and the photoconductivity spectrum, as well as
optical deep level transient spectroscopy \cite{chantre81},
could yield valuable insight into the physics of FEPPC.

 Our
results in GaAs are qualitatively similar to those in AlAs: in
GaAs QWs bounded by Al$_{0.39}$Ga$_{0.61}$As barriers, we
obtained a density increase of about $5 \times 10^{11}$
cm$^{-2}$ after illumination with $V_{BG} = 6$ V (150 V/cm),
while a much smaller density increase ($\sim 1 \times 10^{11}$
cm$^{-2}$) was observed under illumination at $V_{BG} = 0$ V. We
did not measure the dependence of FEPPC on $x_B$ in GaAs QWs.

Lastly, we point out that when both front and back gates are added to the sample, densities as high as $n = 9.6
\times 10^{11}$ cm$^{-2}$ can be obtained after sample illumination. FEPPC may thus also occur in the top AlGaAs
barrier, although the effect is complicated by the presence of intentional Si dopants near the surface of our
sample. We have not studied the front-gate-dependent PPC in greater detail. It is also worth noting that when Si
dopants are present in the {\it back} barrier of the sample, FEPPC seems to have a weaker effect on 2D carrier
density, possibly because the dopant layer partly screens $V_{BG}$ during illumination.

In summary, we describe in this paper the basic properties of a new form of persistent photoconductivity, tunable
with an electric field. By illuminating AlAs or GaAs quantum wells (surrounded Al$_x$Ga$_{1-x}$As barriers) while
applying an electric field ranging from 0 to $\sim 500$ V/cm across the back AlGaAs barrier, we can vary the 2D
electron density from $\sim 0$ to more than $5 \times 10^{11}$ cm$^{-2}$. FEPPC is strongly sensitive to the Al
mole fraction of the back Al$_{x}$Ga$_{1-x}$As barrier: the light-induced 2D density increases with $x$, and
reaches a maximum at $x \simeq 0.4$. The fact that $n$ depends chiefly on the Al content of the back barrier
indicates that the charged centers responsible for back-gate-controlled FEPPC in our samples are mostly located
within that barrier.

The authors gratefully acknowledge useful discussions with D.\ C.\ Tsui. We also thank Audrey Lee and Troy Abe for
help with optical measurements. This work was supported by the NSF.

\begin{center}
{\bf References and Notes}
\end{center}

\break


\begin{references}

\bibitem{lang79}
D.\ V.\ Lang, R.\ A.\ Logan, and M.\ Jaros, Phys.\ Rev.\ B {\bf 19}, 1015 (1979).

\bibitem{burkey76}
B.\ C.\ Burkey, R.\ P.\ Khosla, J.\ R.\ Fischer, and D.\ L.\ Losee, J.\ Appl.\ Phys.\ {\bf 47}, 1095 (1976).

\bibitem{craford68}
M.\ G.\ Craford, G.\ E.\ Stillman, J.\ A.\ Rossi, and N.\ Holonyak, Jr., Phys.\ Rev.\ {\bf 168}, 867 (1968).

\bibitem{zeisel00}
R.\ Zeisel {\it et al.}, Phys.\ Rev.\ B {\bf 61}, R16283 (2000).


\bibitem{chadi89}
D.\ J.\ Chadi and K.\ J.\ Chang, Phys.\ Rev.\ B {\bf 39}, 10063 (1989).

\bibitem{theis86}
T.\ N.\ Theis and S.\ L.\ Wright, Appl.\ Phys.\ Lett.\ {\bf 48}, 1374 (1986).

\bibitem{reddy98}
C.\ V.\ Reddy, K.\ Balakrishnan, H.\ Okumura, and S.\ Yoshida, Appl.\ Phys.\ Lett.\ {\bf 73}, 244 (1998).

\bibitem{chen98}
C.\ Y.\ Chen, T.\ Thio, K.\ L.\ Wang, K.\ W.\ Alt, and P.\ C.\ Sharma, Appl.\ Phys.\ Lett.\ {\bf 73}, 3235 (1998).

\bibitem{zervos99}
M.\ Zervos, M.\ Elliott, and D.\ I.\ Westwood, Appl.\ Phys.\ Lett.\ {\bf 74}, 2026 (1999).

\bibitem{chou85}
M.\ J.\ Chou, D.\ C.\ Tsui, and G.\ Weimann, Appl.\ Phys.\ Lett.\ {\bf 47}, 609 (1985).

\bibitem{pettersson93}
H.\ Pettersson {\it et al.}, J.\ Appl.\ Phys.\ {\bf 74}, 5596 (1993) and references therein.

\bibitem{depoortere02}
E.\ P.\ De Poortere {\it et al.}, Appl.\ Phys.\ Lett.\ {\bf 80}, 1583 (2002).

\bibitem{initialR}
To obtain the initial state in this experiment, we illuminated the sample at $V_{BG} = 4$ V.

\bibitem{BGleak}
A small leakage current ($I_{leak}$) flows between 2D electrons and back gate when the LED is on. In our
experiment, $I_{leak} \sim 1.5$ nA at $V_{BG} = 15$ V, which may thus lower the effective back gate bias applied
to the sample, since $I_{leak}$ causes a voltage drop through the semi-insulating GaAs substrate. Our setup does
not allow measurement of the voltage potential drop across the AlGaAs barrier only.

\bibitem{longillum}
Densities up to twice higher than the nominal $n_{max}$ in AlAs QWs with ``low-x'' Al$_{x}$Ga$_{1-x}$As barriers
($x \lesssim 0.3$) can be obtained by illuminating the sample for much longer times (minutes to tens of minutes).

\bibitem{GaAsPPC}
FEPPC is unlikely to be caused solely by a process within the AlAs QW, since the effect is also observed in GaAs
QWs.

\bibitem{mooney90}
P.\ M.\ Mooney, J.\ Appl.\ Phys.\ {\bf 67}, R1 (1990).

\bibitem{chantre81}
A.\ Chantre, G.\ Vincent, and D.\ Bois, Phys.\ Rev.\ B {\bf 23}, 5335 (1981).

\end{references}
\end{document}